\RequirePackage{fix-cm}
\documentclass[smallextended]{svjour3} 

\smartqed  
\usepackage{amsfonts}
\usepackage{amssymb}
\usepackage{latexsym}
\usepackage{enumerate}
\usepackage{amsmath}
\usepackage{booktabs}
\usepackage{multirow}
\usepackage{bbm}
\usepackage{subfloat}
\usepackage{subfig}
\usepackage{float}
\usepackage[export]{adjustbox}
\usepackage{color}
\usepackage{graphicx}
\usepackage{mathptmx}      
\usepackage{bm}

\usepackage{anysize} 
\usepackage[draft]{changes}

\begin{document}
	
	\title{Smart network based portfolios}
	
	\author{
}

\institute{}
	\author{Gian Paolo Clemente     \and
		Rosanna Grassi      \and 
		Asmerilda Hitaj
	}
	
	
	\institute{
	G. P. Clemente \at
		Catholic University of Milan, Department of Mathematics, Finance and Econometrics. 
		\email{gianpaolo.clemente@unicatt.it}           
		R. Grassi   \at
		University of Milano - Bicocca, Department of Statistics and Quantitative Methods. 
		\email{rosanna.grassi@unimib.it}
		\and
		A. Hitaj, corresponding author,     \at
		University of Milano - Bicocca, Department of Statistics and Quantitative Methods. 
		\email{asmerilda.hitaj1@unimib.it}           
	}
	
	\date{Received: date / Accepted: date}

\maketitle

\abstract{
In this article we deal with the problem of portfolio allocation by enhancing network theory tools.
We use the dependence structure of the correlations network in constructing 
some well-known risk-based models in which the estimation of correlation matrix is a building block in the portfolio optimization. 
We formulate and solve all these portfolio allocation problems using both the standard approach and the network-based approach.  Moreover, in constructing the network-based portfolios we propose the use of two different estimators for the covariance matrix: the sample estimator and the shrinkage toward constant correlation one. All the strategies under analysis are implemented on two high-dimensional portfolios having different characteristics, covering the period from January $2001$ to December $2017$. 
We find that the network-based portfolio consistently better performs and has lower risk compared to the corresponding standard portfolio in an out-of-sample perspective.     

 }
\keywords{Portfolio optimization, Mean-variance, Smart Beta strategies, Networks, Dependence, Interconnectedness}

\section{Introduction}

Modern portfolio theory originates with the seminal work of Markowitz \cite{art:Markowitz1}. This work proposes the innovative idea of relating the return of an asset (the mean) and its risk (the variance) together with those of the other assets in the portfolio selection, through the mean-variance model.
Nevertheless the prominent role in modern investment theory, this model, when applied in asset management setting, can lead to a poor out-of-sample portfolio performance, due to the estimation errors of the input parameters (see, among others, \cite{merton80,jobson1981putting}). Furthermore the risk, measured through the variance and the correlation, is based on expected values representing only a statistical statement about the future. Such measures often cannot capture the true statistical features of the risk and return which often follow highly skewed distributions. \\
To overcome these main drawbacks, several variations and extensions of the original methodology have been proposed in the literature. In \cite{jagannathan2003risk}, a higher out-of-sample performance is derived by imposing specific constraints, and these results have been further confirmed in \cite{behr2013portfolio,hitajBeta}. 
Alternative approaches deal with the problem of optimal portfolio choice by employing a Bayesian methodology to estimate unknown mean-variance parameters reducing the estimation errors. In this context, one of the most prominent is the Bayes-Stein approach based on the idea of shrinkage estimation (\cite{jorion85,jorion86,bauder2018}). The authors in \cite{ledoit03} propose the shrinkage estimator toward the constant correlation, while in \cite{martellini10} this approach has been extended to higher moments such as skewness and kurtosis. Empirical analyses have shown that the use of shrinkage estimators for the mean-variance parameters  often improves the out-of-sample performance (see \cite{hitajBeta,hitaj2018portfolio}).  

It is well known that the effects of the estimation errors of the returns are higher than the effects of the estimation errors of the covariance matrix (see among others \cite{chopra2013effect}).
For this reason many portfolio strategies proposed in literature have put aside returns. 
These are called \textit{risk based} strategies because they rely only on the estimation of the covariance matrix. Some well-known risk based strategies are Global Minimum Variance, Equally Weighted, 
(\cite{DeMiguel07}), Equal Risk Contribution (\cite{Qian,Maillard}) and Maximum Diversified Portfolio (\cite{Choueifaty}). 
The \textit{risk based} strategies are also called \textit{Smart Beta}\footnote{ The term \textsl{Smart Beta} is popular to denote any strategy which attempts to take advantage of the benefits of traditional passive investments, adding a source of outperformance in order to beat traditional market capitalization-weighted indices. For more information on the \textsl{risk-based} strategies see e.g. \cite{amenc2013} and the references therein.} strategies, as they are also proposed as alternatives to market capitalization-weighted indices, which are claimed to be not efficient, see \cite{Choueifaty}. 
\color{black} 


In the last few years the problem of asset allocation has been discussed under a different perspective, using network theory to represent the financial market. Indeed, in network-based portfolio models, the correlation matrix is included in the network structure, in order to reproduce the dependence among the assets (see, for instance, \cite{Mantegna99,Onnela2003,pozzi13,Zhan2015}), providing in this way useful insights in the portfolio selection process.
In particular, the minimum spanning tree has been used in \cite{Onnela2003}, the authors in \cite{pozzi13} use Planar Maximally Filtered Graphs, while in \cite{Zhan2015} hierarchical clustering trees and neighbour-nets have been applied in order to reduce the complexity of the network, characterizing the heterogeneous spreading of risk across a financial market. The work of Peralta and Zareei (\cite{peralta16}) establishes a bridge between Markowitz's framework and the network theory, showing a negative relationship between optimal portfolio weights and the centrality of assets in the financial market network. As a result, the  centrality  measures  of constructed  networks  can  be  used  to facilitate the portfolio selection. A generalization to this approach has been provided in \cite{Vyrost2019}.\\
Recently, an alternative methodology to tackle the asset allocation problem using the network theory has been proposed in \cite{ANORclegrahit}. Specifically, the authors catch how much a node is embedded in the system, by adapting to this context the clustering coefficient, a specific network index (see \cite{Barrat_2004,CleGra,Fagiolo_2007,mcassey2015clustering,WasFaust,watts1998collective}), meaningful in financial literature to assess systemic risk \cite{Bongini,Minoiu,Tabak}. The underlying structure of the financial market network is used as an effective tool in enhancing the portfolio selection process. In particular, the optimal allocation is obtained by maximizing a specific objective function that takes into account the interconnectedness of the system, unlike the classical global minimum variance model that is based only on the pairwise correlation between assets. Furthermore, in constructing the dependence structure of the portfolio network, various dependence measures are tested, namely, the  Pearson correlation, Kendall correlation and lower tail dependence. All these measures are estimated using the sample approach. The results obtained  in \cite{ANORclegrahit} show that, independently from the length of the rolling window and from the used dependence structure, the network-based portfolio leads to better out-of-sample performance compared with the classical approach.\\ 
The aim of this paper is to move one step further by enhancing the role of the network theory in solving portfolio allocation problem. This paper 
contributes to the existing literature along various dimensions. \\ First, we exploit the network theory also for constructing the 
Smart Beta strategies and the mean-variance portfolio, where different values of trade-off parameter are considered. 
Second, we evaluate the shrinkage estimator, toward the Constant Correlation, also in the network-based portfolio strategies. Third we empirically test the out-of-sample performance of the proposed methodology, shedding some light on the network-based portfolio strategies. \\
Specifically, the Pearson correlation is used in order to capture the dependence structure of the portfolio network. Moreover, we apply the network theory to various well-known models in which the estimation of the correlation matrix is a building block in the portfolio optimization. 
We consider in this paper Equally Risk Contribution, Maximum Diversified Portfolio, Global Minimum Variance, and the mean-variance model. In this last case we consider different levels of the trade-off parameter. 
Moreover, since recent academic papers and practitioner publications suggest that equal-weighted portfolios appear to outperform various other price-weighted or value-weighted strategies (see, e.g., \cite{DeMiguel07}), we also include the Equally Weighted (EW) portfolio in our analysis.
\color{black}
Through empirical analyses, we test the impact of the estimation method on both the \textsl{standard} and \textsl{network-based} portfolios. For the sake of completeness, two different high-dimensional portfolios with different characteristics are considered. The first portfolio is composed by 266 among largest banks and insurance companies in the world. The second portfolio is composed by the components of the S\&P 100 index. Both datasets contain daily returns in the time-period ranging from January 2001 to December 2017. All the obtained portfolios are compared in an out-of-sample perspective using some well-known performance measures. Main results show that, in the majority of the cases, the use of  network-based approach leads to higher out-of-sample performances and lower volatility with respect to the corresponding \textsl{sample} strategy.  The network-based portfolio is more robust with respect to the standard approach being only slightly affected by the estimation method of the covariance matrix. The out of-sample results suggest that the network-based strategy represents a viable alternative to classical portfolio strategies.\\
The remainder of the paper is organized as follows. Section \ref{par:methodology} briefly recalls the investor's problems for each strategy under analysis. Section \ref{estMeth} explains the two estimation methods used for the covariance matrix. Section \ref{sec:corrnet} explains in detail the approach of portfolio selection via network theory. Section \ref{result} presents the empirical analysis and Section \ref{concl} draws some conclusions.

\section {Portfolio selection strategies} 
\label{par:methodology}
In this section, we briefly set out the strategies used in the rest of the paper for the empirical analysis. We first introduce what we refer to as standard strategies. We start with the mean-variance problem and then we describe the most important Smart Beta approaches proposed as alternatives to the  market capitalization-weighted indices, in the equity world. 

\subsubsection*{Mean-variance (MV)} 
\label{estMeth}
To fix the notation, let us first introduce the standard \textsl{mean-variance} model for a portfolio with $N$ risky assets. Let $R_i$ be the random variable (r.v.) of daily log returns. Let $\textbf{r}=[r_i]_{i=1,\ldots,N}$  be the returns' vector observed in a specific time period/window ($w$) and $\bm{\mu}$ ($\bm{\Sigma}$) be the mean vector (variance-covariance matrix) between assets estimated in the same period. Let $\mathbf{e}$ and  $\mathbf{x}=[x_i]_{i=1,\ldots,N}$ be, respectively, the vector of ones and the vector of portfolio  weights, i.e. the proportional investments in the $N$ risky assets. We denote with $\mu_p=\textbf{x}^{T} \bm{\mu}$,  $\sigma_i$ and $\sigma_P={\sqrt{\mathbf{x}^{T}\bm{\Sigma}\ \mathbf{x}}}$, respectively, the portfolio mean, the standard deviation of the $i^{th}$ asset and  the standard deviation of the portfolio. We recall that all optimization models, considered in this paper, include realistic investment constraints such as budged constraint, $\mathbf{e}^T\mathbf{x}=1$, and non-short selling constraints $x_i \geq 0$ $\forall x_i$ since many institutional investors are restricted to long positions only. 

The mean-variance model proposed in \cite{art:Markowitz1} consists in optimizing a trade-off between risk and return. The standard mean-variance (MV) portfolio optimization problem is given by:
\begin{equation}
\label{eq:MV}
\left\{
\begin{array}{ll}
\smallskip
\min\limits_{\mathbf{x}} \ \ \ \lambda \textbf{x}^{T}\bm{\Sigma}\textbf{x} - (1-\lambda)\textbf{x}^{T} \bm{\mu} \\ \smallskip
\mathbf{e}^T\mathbf{x}=1\\
0 \leq x_i \leq 1, & \hbox{$i=1, \dots,N$}
\end{array},
\right.
\end{equation}
where $\lambda \in [0,\ 1]$ expresses the trade-off between risk and return of the portfolio.  It is possible to compute alternative points on the efficient frontier by solving problem (\ref{eq:MV}) for different levels of $\lambda$.

It is evident from \eqref{eq:MV} that the  MV portfolio optimization relies on estimators of the means and covariances of the asset returns. This means that the MV portfolio strongly depends on the input data, see among others \cite{jorion1992portfolio} and \cite{chopra2013effect}. The estimation methods used in this paper will be explained in more details in Subsection \ref{estMeth}

\subsubsection*{Global Minimum-Variance portfolio (GMV)} 
The GMV strategy select portfolio weights that minimize the variance of the portfolio  ignoring completely the portfolio return. The GMV optimization problem is formulated as:
\begin{equation}
\label{Classoptimalconst}
\left\{
\begin{array}{ll}
\smallskip
\min\limits_{\mathbf{x}} \ \ \ \textbf{x}^{T}\bm{\Sigma}\textbf{x}\\ \smallskip
\mathbf{e}^T\mathbf{x}=1\\
0 \leq x_i \leq 1, & \hbox{$i=1, \dots,N$}
\end{array},
\right.
\end{equation}

\subsubsection*{Equally Weighted portfolio (EW)}
The EW strategy consists in holding a portfolio characterized by the same weight $\frac{1}{N}$ in each component. In the literature, it has been empirically showed, see \cite{DeMiguel07}, that the EW portfolios perform better than many other quantitative models, with higher Sharpe Ratio and Certainty Equivalent return. Being the weights equally allocated among the assets, this strategy disregards the data and of course it does not require any optimization or estimation procedure.  

\subsubsection*{Equal Risk Contribution portfolio (ERC)} 
The equal risk contribution strategy (ERC) is characterized by weights such that each asset provides the same contribution to the risk of the portfolio. 
\newline The marginal contribution of asset $i$ to the risk of the portfolio is:

\begin{equation*}
\partial_{x_{i}}\sigma_{P}=\frac{\partial \sigma_{P}}{\partial x_{i}}=\frac{\left(\bm{\Sigma}\ \textbf{x}\right)_{i}}{\sqrt{\mathbf{x}^{T}\bm{\Sigma}\ \mathbf{x}}}.
\end{equation*}

\noindent
Hence, $\sigma_{i}(x)= x_{i} \partial_{x_{i}}\sigma_{P}$ represents the risk contribution of the $i^{th}$ asset to the portfolio $P$. 
The authors in \cite {Maillard} have demonstrated  that the portfolio risk can be expressed as: $$\sigma_{P}=\sum_{i}^{N} \sigma_{i} (x)$$ that is the sum of risk contributions of the assets. 
The characterizing property of the ERC strategy is that weights are such that $\sigma_{i} (x)=\sigma_{j}(x)$ $\forall \ i,j$. The result is a portfolio extremely diversified in terms of risk.
To obtain the optimal weights we have to solve an optimization problem consisting in minimizing the sum of all squared deviations under budged and non short-selling constraints. The mathematical formulation is the following: 
\begin{equation}
\left\{
\begin{array}
[c]{l}%
\smallskip
\underset{\textbf{x}}{\min}\ \sum_{i=1}^N \ \sum_{j=1}^N \ \left(x_{i}(\bm{\Sigma}\ \textbf{x})_{i}-x_{j}(\bm{\Sigma}\ \textbf{x})_{j}\right)^2\\
\mathbf{e}^T\mathbf{x}=1 \\
x_i \geq 0 \ \  i=1,...N
\end{array}
\right.,
\label{erc}
\end{equation}

\subsubsection*{Maximum Diversified portfolio (MDP)}
The Maximum Diversification approach aims to construct a portfolio that maximizes the benefits from diversification. This goal can be achieved by solving a maximization problem where the objective function is given by the so-called Diversification Ratio $DR=\frac{\sum_{i=1}^N{x_{i}\sigma_{i}}}{\sigma_P}$,  under the usual constraints. The mathematical formulation for the  MDP strategy is:
\begin{equation}
\left\{
\begin{array}
[c]{l}%
\smallskip
\underset{\mathbf{x}}{\max}\ \frac{\sum_{i=1}^N{x_{i}\sigma_{i}}}{\sqrt{\mathbf{x}^{T} \Sigma \mathbf{x}}}\\
\mathbf{e}^T\mathbf{x}=1 \\
x_i \geq 0 \ \  i=1,...N
\end{array}
\right.
\label{mdp}
\end{equation}

\noindent
This approach creates portfolios characterized by minimally correlated assets, providing lower risk levels and higher returns than market cap-weighted portfolios strategies (see \cite{Choueifaty}). 

\subsection{Estimation methods for covariance matrix}
\label{estMeth}
Unlike the investment problems \eqref{Classoptimalconst},  \eqref{erc} and \eqref{mdp}, in which only the estimate of the covariance matrix between assets in a given time interval is needed, we have to estimate both the covariance matrix and the mean vector in order to solve problem \eqref{eq:MV}. A common way to estimate them is through the \emph{sample approach}.  
This method allows to obtain each component $\hat{\mu}_i$ and $\hat{\sigma}_{i,j}$ by means of classical unbiased estimators.  
However, it is well known that the sample estimator of historical returns is likely to generate high sampling error. For this reason several methods have been introduced in order to improve the estimation of moments and comoments, see among others \cite{jorion85}, \cite{jorion86}, \cite{ledoit03} and \cite{martellini10}. 
These methods are grounded on the idea of imposing some structure on the moments (comoments) with the aim to reduce the number of parameters, leading in this way to a reduction of the sampling error at the cost of specification error. 

\noindent It is worth stressing that in this work we mainly focus on the estimation of the covariance matrix. For this reason, we estimate the mean only through the sample approach, whereas we pay more attention to the estimate of the covariance matrix, considering also the shrinkage toward the constant correlation (CC) method (see \cite{EdwinGruber,ledoit03}). The idea of the CC approach is to estimate the covariance based on the fact that the correlation is assumed constant for each pair of assets, and it is given by the average of all the sample correlation coefficients (see \cite{EdwinGruber}). The covariance between two assets is then computed as 
\begin{equation}
\sigma_{i,j}^{CC}=\hat{\sigma}_{i}\hat{\sigma}_{j}\frac{1}{N(N-1)} \sum_{\substack{{i,j=1}\\i \neq j}}^N{\left(\hat{\rho}_{i,j}\right)},
\end{equation}
where $\hat{\rho}_{i,j}$ is the sample correlation between asset $i$ and $j$. \\
This approach resizes the problem, as only one correlation coefficient and $N$ standard deviations have to be estimated. The $\bm{\Sigma^{CC}}$ covariance matrix, constructed by using previous formula, is characterized by a lower estimation risk due to the assumed structure, nevertheless it involves some misspecification in the artificial structure imposed by this estimator. 
In the attempt to find a trade-off between the sample risk and the model risk, the authors in \cite{ledoit03} introduce the asymptotically optimal linear combination of the sample estimator and the structured estimator (in our case, the CC estimator) in the context of the covariance matrix, with the weight given by the optimal shrinkage intensity $\kappa$\footnote{For further information on how is estimated the shrinkage intensity $\kappa$ see \cite{ledoit03}.}. Therefore, the shrinkage toward CC covariance matrix is given by:
\begin{equation}
\sigma_{i,j}^{shrink}=\kappa \sigma_{i,j}^{CC}+ (1-\kappa) \hat{\sigma}_{i,j}.
\label{eq:corr_shrink}
\end{equation}

In the following section we recall the network correlation-based portfolio model and  explain both estimators of the variance-covariance matrix (sample and shrinkage toward CC) can be used.


\section{Optimal portfolio via network theory}\label{sec:corrnet}

The portfolio selection problem and its variants can be formulated in a networks context and several researchers dealt with the assets allocation problem using network theory tools, contributing to the related literature (\cite{peralta16,ANORclegrahit,Li2019,pozzi13}).
All these articles share the same framework, namely the financial market is represented as a network, in which nodes are assets and weights on the edges identify a dependence measure between returns.\\
We describe in this Section the approach proposed by \cite{ANORclegrahit}. The authors  formulate an investment strategy that benefits from the knowledge of the dependency structure that characterize the market. Unlike the risk-based strategies, based on an objective function that accounts for pairwise correlations among assets, the objective function considers here the interconnectedness of the whole system. 

In order to make the paper self-consistent, we briefly remind some preliminary definitions and notations about networks. A network $G=(V,E)$ consists in a set $V$ of nodes and a set $E$ of edges between nodes, where the edge $(i,j)$ connects a pair of nodes $i$ and $j$.
If $(j,i)\in E$ whenever $(i,j)\in E$, the network is undirected. A network is complete if every pair of vertices is connected by an edge.
We denote with $\mathbf{A}$ the real $N$-square matrix whose elements are $a_{ij}=1$ whenever $(i,j)\in E$ and $0$ otherwise (the adjacency matrix). 
A network is weighted if a weight $w_{ij} \in \mathbb{R}$ is associated to each edge $(i,j)$. In this case, both adjacency relationships between vertices of $G$ and weights on the edges are described by a non negative, real $N$-square matrix $\textbf{W}$ (the weighted adjacency matrix).
We denote with $k_{i}$ and $s_{i}$ the degree and strength of the node $i$ $(i=1,...,N)$, respectively.

Relationships between assets are quantified through three different levels of dependence. For a sake of brevity, we report here only the approach referring to the classical linear correlation network, which is the most used dependence measure in the literature. 
Since all assets are correlated in the market, the correlation structure is represented through a weighted, complete and undirected network $G$, where weights on the edges are given by the Pearson correlation coefficient between them, that is $w_{ij}=\rho\left(R_{i},R_{j}\right)$ $\forall i \neq j$. In order to assure nonnegative weights, a distance can be associated with the correlation coefficient (see \cite{giudici2016,Mantegna99,onnela03}). In our case, this transformation does not affect the results in terms of optimal portfolio.


The extension of the pairwise correlations, included in the quadratic form of the problem (\ref{eq:MV}), to a general intercorrelation among all stocks at the same time is obtained optimizing a function that includes the clustering coefficient. The classic clustering coefficient and its variants defined in the literature (see \cite{watts1998collective,Roy,Fagiolo_2007,CleGra}) are not computable for complete networks, then we have to adapt its formulation to this framework.

Following a similar procedure to that proposed by \cite{mcassey2015clustering}, a threshold $s\in[-1,1]$ is introduced on the matrix $\mathbf{W}$ in order to define the new matrix $\mathbf{A}_{s}$, whose elements $a_{ij}^s$ are 
\begin{equation}
a_{ij}^s = \begin{cases} 1 & \mbox{if }  w_{ij} \geq s \\ 0 & \mbox{otherwise}
\end{cases}.
\end{equation}

\color{black}

\noindent 
$\mathbf{A}_s$ is the adjacency matrix describing the existing links in the network with weights $w_{ij}$ at, or above the threshold $s$. 
Through this matrix we are selecting the strongest edges, namely those greater than a given threshold, bringing out the mean cluster prevalence of the network.
On this new network we compute the clustering coefficient proposed in \cite{watts1998collective} and then we repeat the process, varying the threshold $s$. 
The clustering coefficient $C_{i}$ for a node $i$ corresponding to
the graph is the average of $C_{i}(\mathbf{A}_{s})$ overall
$s\in[-1,1]$:
\begin{equation}\label{clust_tot}
C_{i}=\int_{-1}^{1}C_{i}(\mathbf{A}_{s})ds
\end{equation}

Since $0\leq C_i\leq1$, $C_{i}$ is well-defined.  
Now, we define the $N$-square matrix $\mathbf{C}$, of entries 
\begin{equation}
c_{ij} = \begin{cases} C_{i}C_{j} & \mbox{if }  i \neq j \\ 1 & \mbox{otherwise}
\end{cases}.
\end{equation}

\noindent This matrix accounts for the level of interconnection for all pairs with the whole system, therefore, it can be used to construct the matrix 
\begin{equation}
\mathbf{H}=\bm{\Delta}^T\mathbf{C}\bm{\Delta}
\label{eq:matrH}
\end{equation} where $\bm{\Delta}=diag(s_{i})$ is a diagonal matrix with diagonal entries $s_i=\frac{\hat{\sigma}_{i}}{\sqrt{\sum_{i=1}^{N}\hat{\sigma}_{i}^{2}}}$. \\
Notice that the element $s_{i}$ considers the contribute of the standard deviation of the returns $i$ with respect to the total standard deviation, computed in case of independence. 
In (\cite{ANORclegrahit}), the authors solve the optimization problem defined in \eqref{Classoptimalconst} replacing the covariance matrix $\bm{\Sigma}$ with $\mathbf{H}$.

The main difference between the classical and the network portfolio selection problem is due to the use of the interconnectedness matrix in order to consider how much each couple of assets is related to the system. In particular, being $\mathbf{C}$ dependent on a network-based measure of systemic risk (i.e. the clustering coefficient), we are implicitly including a measure of the state of stress of the financial system in the time period.

\section{Dataset description and empirical analysis} \label{result}

\subsection{Dataset description}
The goal of this section is to examine the out-of-sample properties of the Smart Beta and mean- variance  network-based portfolios in which the covariance matrix is estimated using the network theory though the methodology described in Section \ref{sec:corrnet}. In particular, we make a comparison between network-based portfolios and standard portfolio strategies, 
where both the sample and the shrinkage estimators of the covariance matrix are considered.
We summarize in Table \ref{tab:Models} the alternative asset allocation models applied in this analysis.

\begin{table}[htbp]
	\centering
	\begin{tabular}{clr}
		\hline
		Number & \multicolumn{1}{c}{Model} & \multicolumn{1}{c|}{Label} \\
		\hline
		1     & Standard sample based mean-variance & S-sample MV \\
		2     & Standard shrinkage toward constant correlation based mean-variance & S-shrinkage MV \\
		3     & Network sample based mean-variance & NB-sample MV \\
		4     & Network shrinkage toward constant correlation based mean-variance & NB-shrinkage MV \\
		\hline
		5     & Standard sample based Maximum Diversified Portfolio (MDP) & S-sample MDP \\
		6     & Standard shrinkage toward constant correlation based MDP & S-shrinkage MDP \\
		7     & Network sample based MDP & NB-sample MDP \\
		8     & Network shrinkage toward constant correlation based MDP & NB-shrinkage MDP \\
		\hline
		9     & Standard sample based Equally Risk Contribution (ERC) & S-sample ERC \\
		10    & Standard shrinkage toward constant correlation based ERC & S-shrinkage ERC \\
		11    & Network sample based ERC & NB-sample ERC \\
		12    & Network shrinkage toward constant correlation based ERC & NB-shrinkage ERC \\
		\hline
		13    & Standard sample based Global Minimum Variance (GMV) & S-sample GMV \\
		14    & Standard shrinkage toward constant correlation based GMV & S-shrinkage GMV \\
		15    & Network sample based GMV & NB-sample GMV \\
		16    & Network shrinkage toward constant correlation based GMV & NB-shrinkage GMV \\
		\midrule
		17    & Equally Weighted & EW \\
		\hline
	\end{tabular}%
	\caption{List of asset allocation models considered in the empirical study. The last column
		of the table indicates the label used to refer to each strategy in the empirical section, where the performance of the various approaches is compared.}
	\label{tab:Models}%
\end{table}%

\noindent
As a robustness check we consider two large-dimensional portfolios with different characteristics. The investment universe of the first portfolio is composed by 266 among largest banks and insurance companies in the world\footnote{ The greatest firms by market capitalization in the banking and insurance sector are considered.}. In particular we have 120 insurers and 144 banks. The investment universe of the second portfolio consists of the components of the S\&P 100 index.  
The datasets of the two portfolios under analysis contain daily returns in the time-period ranging from January 2001 to December 2017\footnote{Data have been downloaded from Bloomberg \cite{bloombe}}.

\noindent
All the portfolios discussed in this paper are analyzed and compared in an out-of-sample perspective. In particular the first four moments, the Sharpe Ratio (SR), the Omega Ratio (OR) the Information Ratio (IR) and the out-of-sample performance are used to compare the portfolios. All these aspects are investigated through a rolling window methodology,  which is characterized by an in-sample period of length $n$ and an out-of-sample period of length $k$. This means that the first in-sample window of width $n$ contains the observations of all the components in the portfolio from $t=1$ to $t=n$. The dataset of the first in-sample window is used to estimate the optimal weights, using the different portfolio selection models considered in this paper and listed in Table \ref{tab:Models}. These optimal weights are then invested in the out-of-sample period, from $t=n+1$ to  $t=n+k$, where the out-of-sample performance is computed.  
The process is repeated rolling the window $k$ steps forward. Hence, weights are updated by solving the optimal allocation problem in the new subsample and the performance is estimated once again using data from $t=n+k+1$ to $n+2k$. Repeating these steps until the end of the dataset is reached, we buy-and-hold the portfolios and we record out-of-sample portfolio returns in each rebalancing period.

To ensure the robustness of our results, we analyze two different estimation window lengths; namely, 6 months in-sample and 1 month out-of-sample and two years in-sample and 1 month out-of-sample.

\subsection{Performance measures}

In order to assess the magnitude of potential gains/losses that can be attained by an investor adopting a network-based portfolio selection, we implement an out-of-sample analysis. For this reason several performance measures are calculated. First, for each optimization strategy, we compute the first four moments of the out-of-sample portfolio returns. Further, for each strategy $j$, we determine the out-of-sample Sharpe Ratio of the optimal portfolio: 
\begin{equation*}
SR_j^{\star}=\frac{\mu_{p_j} ^{\star}-\mu_f}{\sigma_{p_j}^{\star}},
\end{equation*}
where $\mu_{p_j}^{\star}$\footnote{In general, the superscript  $\star$ indicates that the statistic is calculated using the out-of-sample time series of returns of the  optimal portfolio.} and $\sigma_{p_j}^{\star}$ are, respectively, the average return and the standard deviation of the optimal portfolio according to the strategy $j$ and $\mu_f$ indicates the average risk-free rate\footnote{As a proxy for the risk-free rate  the literature suggests the use of \textsl{1 month} or \textsl{3 months} maturity U.S. Treasury Bills (see e.g \cite{MaDeMe}) or, alternatively, an exogenously given value (for instance,  $\mu_f=5\%$ is considered in \cite{Brennan}).  In the empirical part of this paper, for illustration purposes, we set  $\mu_f=0 \%$.}.
This ratio measures the average return of a portfolio in excess of the risk-free rate, also called the risk premium, as a fraction of the portfolio total risk, measured by its standard deviation. 
As alternatives performance measures we also calculate the Information ratio (IR) and the  Omega ratio (OR). The Information ratio of the optimal portfolio is defined as:
\begin{equation*}
IR_j^{\star}=\frac{\mu_{p_j} ^{\star}-\mu_{p_{ref}} ^{\star}}{\sigma(r_{p,j}^{\star}-r_{p_{ref}}^{\star})}.
\end{equation*}
where $\mu_{p_{ref}} ^{\star}$ is the average return of the reference portfolio and $r_{p,j}^{\star},r_{p_{ref}}^{\star}$ represent the out-of-sample time series of optimal
portfolio returns corresponding to a strategy $j$ and the reference strategy, respectively. 

Once identified the reference portfolio, managers seek to maximize $IR_{j}$, i.e. to reconcile a high residual return and a low tracking error. This ratio allows to check if the risk taken by the manager in deviating from the reference portfolio is sufficiently rewarded. \\
The \textsl{Omega Ratio} has been introduced by Keating and Shadwick in \cite{keating2002universal} and it is defined as:
\begin{equation*}
OR_j^{\star} = \frac{\int_{\epsilon}^{+\infty} (1-F_j(x))\,dx}{\int_{-\infty}^{\epsilon} F_j(x)dx}=\frac{\mathbb{E}\left(r^{\star}_{p_j}-\epsilon\right)^+}{\mathbb{E}\left(\epsilon-r^{\star}_{p_j}\right)^+},
\end{equation*}
where $F_j(x)$ is the cumulative distribution function of the portfolio returns for a strategy $j$  
and $\epsilon$ is a specified threshold\footnote{We point out that $OR$ ratio is very sensitive to values of $\epsilon$ which can be different from investor to investor. In the empirical analysis $\epsilon$ is set equal to $0$.}. 
\color{black} Returns below (respectively above) the threshold are considered as losses (respectively gains). In general, a value of the $OR_{j}$ greater than 1 indicates that  strategy $j$ provides more expected gains than expected losses. The portfolio with the highest ratio will be preferred by the investor. The $OR_{j}$ implicitly embodies all the moments of the return distribution without any a-priori assumption. 

\subsection{Empirical results}
\begin{figure}[H]
	\centering
	\subfloat[Pearson Correlation Network computed by using returns of Banks and Insurers dataset (based on sample estimation) referred to the last window, from the beginning of December 2015 to the end of November 2017. Bullets size is proportional to the standard deviation of each firm. Edges opacity is proportional to edges weights (i.e. intensity of correlations).]{\fbox{\includegraphics[width=0.6\linewidth]{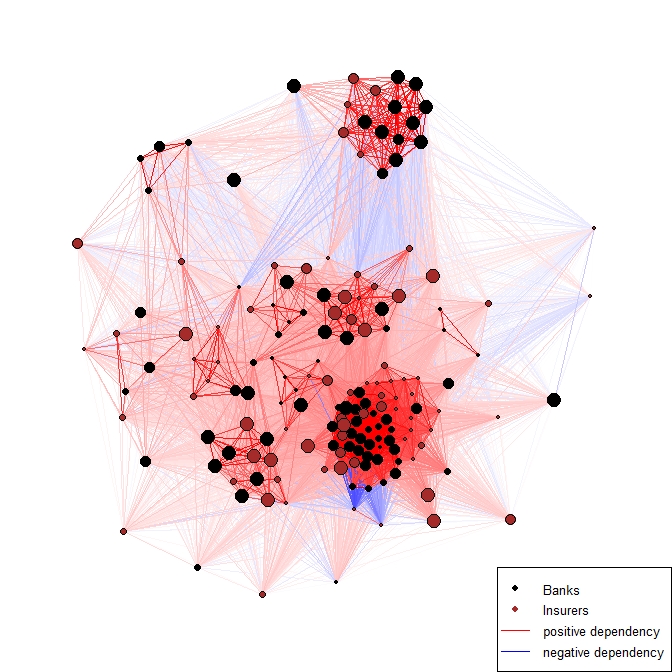}}\label{fig:pearsCorr}}\ \hspace{0.5mm}
	\subfloat[The optimal network-based sample GMV portfolio referred to the same period as in \ref{fig:pearsCorr}, where the covariance matrix is estimated using the sample approach (NB-sample GMV). In this figure, the bullets size is proportional to the allocated weight. Edges opacity is proportional to edges weights.]{\fbox{\includegraphics[width=0.6\linewidth]{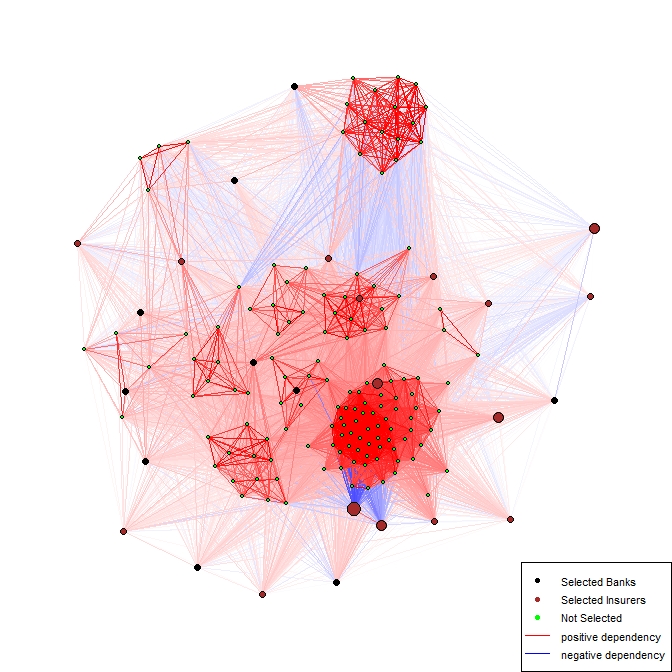}}\label{fig:GMVNet}}
	\caption{}
	\label{F:BI2Ynet}	
\end{figure}

As previously explained, for robustness purpose we consider two large-dimensional portfolios. The first is the Banks and Insurers portfolio composed by 266 among largest banks and insurance companies in the world. The second portfolio is composed by the assets of the S\&P 100 index. This empirical analysis is based on a buy and hold strategy. For the sake of completeness we consider two 
strategies, with in-sample period of two years and six months, respectively, and out-of-sample period of one month.
For the sake of brevity, we report in the following the results obtained for the Banks and Insurers portfolio with a rolling window of two years in-sample and one month out-of-sample. All the other results are reported in the supplementary material.  
We focus on this specific dataset because it is rarely used in the empirical analysis and through network visualization it is easier to capture the dependence structure between banks and insurers. 
Moreover, we can have a clear and informative drawing of the portfolio composition. \\
To have a preliminary idea of the portfolio we depict in Figure \ref{fig:pearsCorr} the sample correlation network obtained in the last window that covers the period December 2015-November 2017. As previously pointed out, each node represents a firm (a bank or an insurer) and the weighted edge $\left(i, j\right)$ measures the correlation between firms $i$ and $j$. \\
As described in Section \ref{sec:corrnet}, a network-based portfolio model is solved, where in the optimal problem the interconnections' structure between assets is caught by means of the clustering coefficient. In particular, the pairwise correlations affect the level of interconnections and, therefore, the optimal solution.  
We report in Figure \ref{fig:GMVNet} the optimal solution of the sample network-based GMV problem, i.e. NB-sample GMV,   for the same window $w$ considered in Figure \ref{fig:pearsCorr}. In this network representation, the size of bullets is instead proportional to the allocated weight. We observe that the initial endowment is invested in only 26 firms, specifically 10 banks and 16 insurance companies. However, approximatively 94\% of the total amount is invested in insurers that are characterized on average in this time period by both a lower volatility and a lower clustering coefficient. \\
It is worth noting the case of two insurers, Nationwide Mutual Insurance Company and One America, which are characterized by the lowest standard deviations and a high proportion of negative pairwise correlations (for instance, approximatively 90\% of correlations between Nationwide Mutual and other firms is lower than zero).
As expected, the optimal portfolio allocates a high proportion of the initial endowment in these two firms (54\% and 17\% respectively).

\begin{figure}[H]
	\centering
	\subfloat[MDP portfolios through standard and network-based methodology using sample and shrinkage estimators.]{\fbox{\includegraphics[ height=5cm, width=7.5cm]{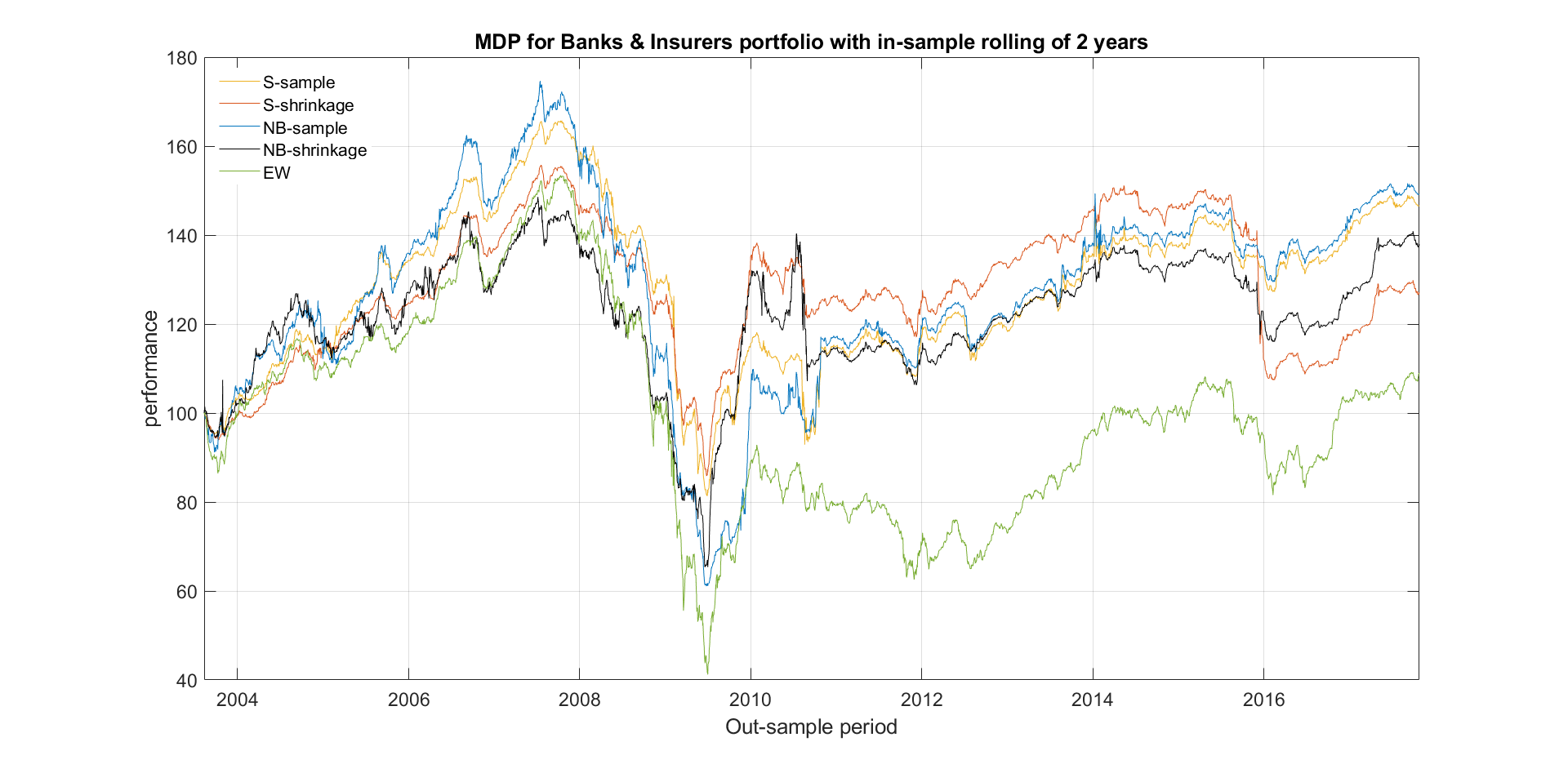}}\label{fig:BI2Y_MDP}}\ \hspace{0.5mm}
	\subfloat[ERC portfolios through standard and network-based methodology using sample and shrinkage estimators.]{\fbox{\includegraphics[ height=5cm, width=7.5cm]{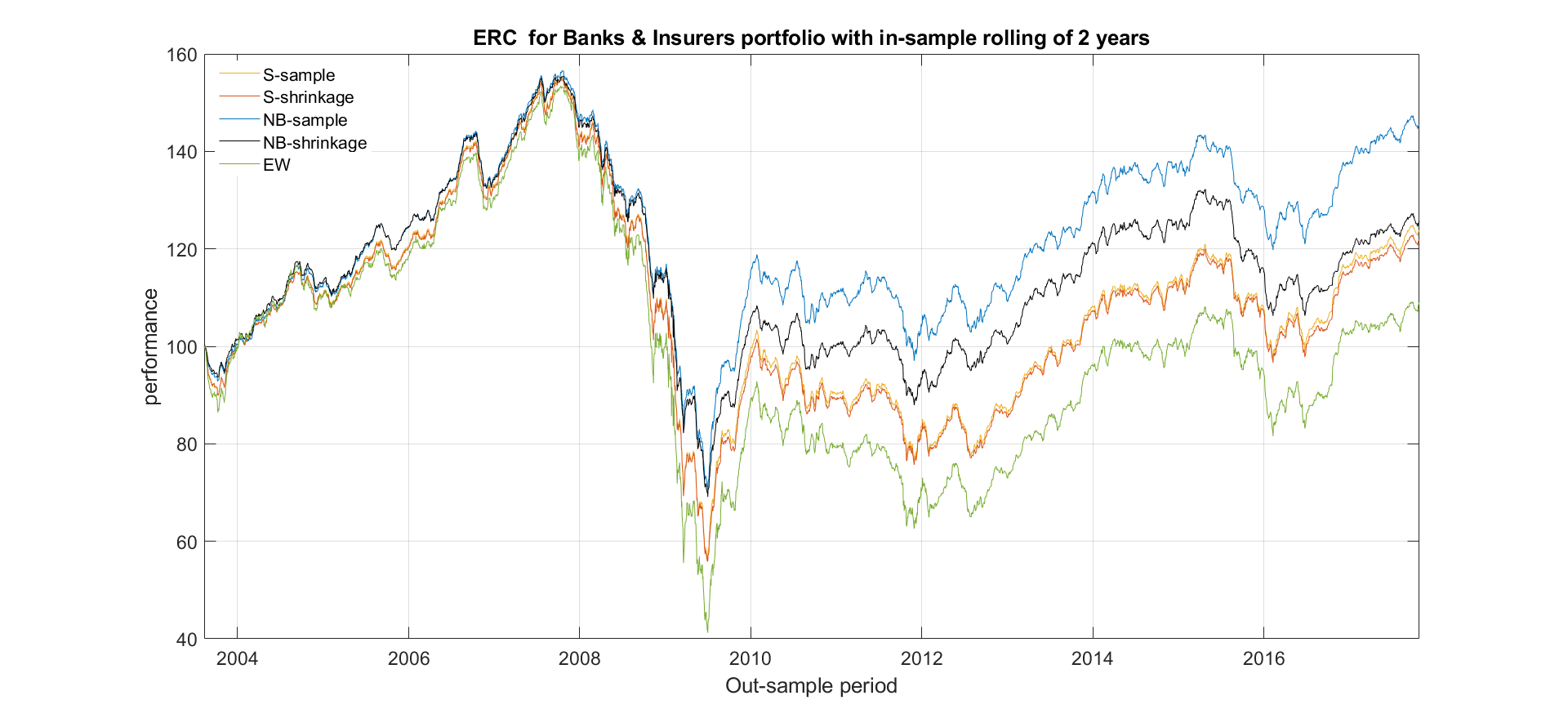}}\label{fig:BI2Y_ERC}}\ 
	\subfloat[GMV portfolios through standard and network-based methodology using sample and shrinkage estimators.]{\fbox{\includegraphics[ height=5cm, width=7.5cm]{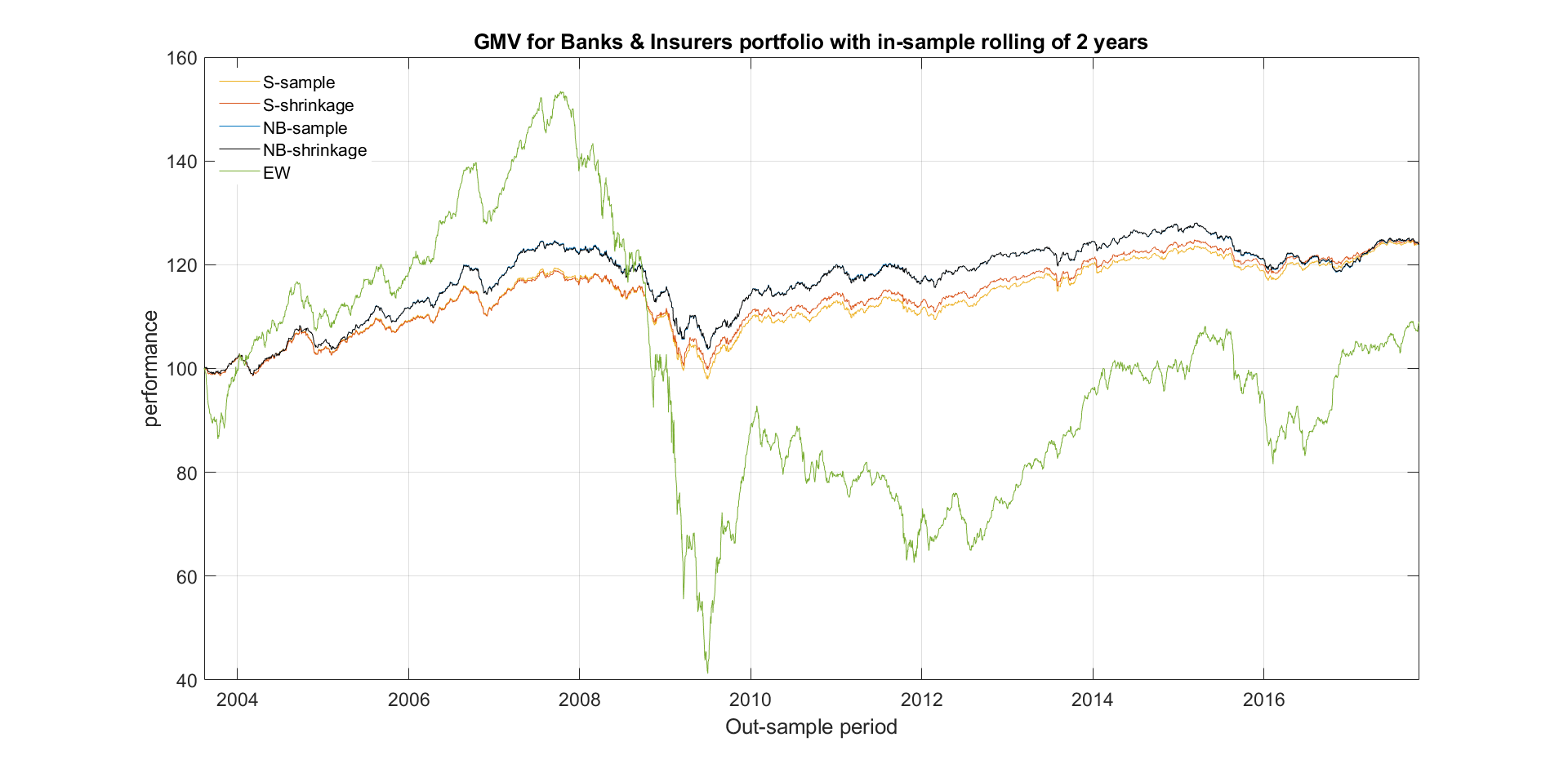}}\label{fig:BI2y_GMV}}\ \hspace{0.5mm}
	\subfloat[Best out-of-sample performances for each smart beta portfolio.]
	{\fbox{\includegraphics[ height=5cm, width=7.5cm]{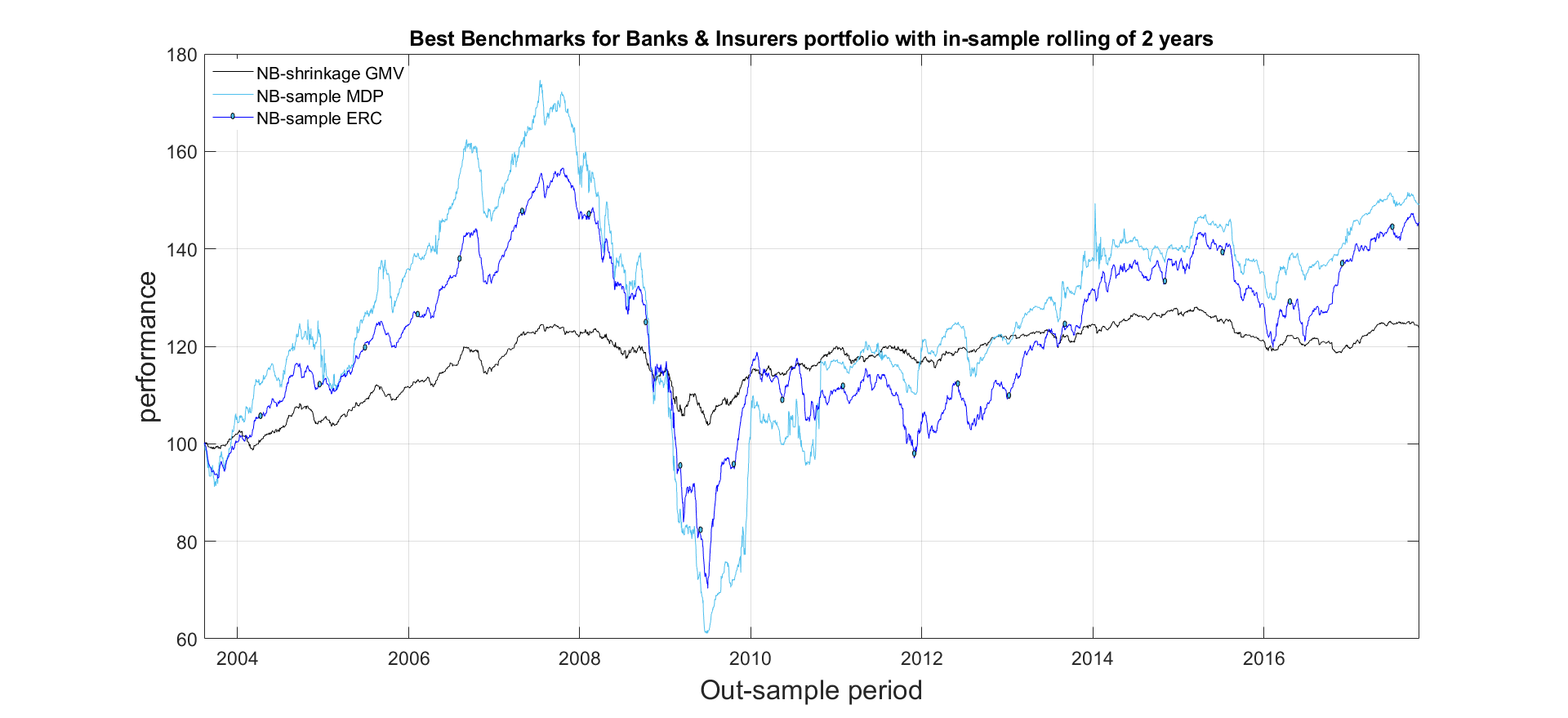}}\label{fig:BI2Y_Best_Bench}}\\
	\caption{Out-of-sample performances for Banks and Insurers dataset with a rolling window of 2 years in-sample and 1 month out-sample. In Figures \ref{fig:BI2Y_MDP}, \ref{fig:BI2Y_ERC} and \ref{fig:BI2y_GMV} the out-of-sample performances of EW, S-sample, S-Shrinkage, NB-sample and NB-Shrinkage of, respectively, MDP, ERC and GMV models are reported. In Figure \ref{fig:BI2Y_Best_Bench} the best out-of-sample performances for each Smart Beta portfolio (MDP, ERC and GMV) are reported.}
	\label{fig:smart_beta_perf}	
\end{figure}

In the following we report the out-of-sample performances of all the models under investigation. We analyse the Smart Beta and the MV optimal portfolios obtained using standard and network-based approaches for both sample and shrinkage estimators.\\ In Figure \ref{fig:smart_beta_perf} the out-of-sample performances of the Smart Beta models under analysis are reported. 


We observe that in all cases the EW strategy has the worst performance. Concerning 
the other strategies, focusing on MDP portfolios  (Figure \ref{fig:BI2Y_MDP}) the sample estimators lead to best performing models over time, with a prevalence of the network-based model (NB-sample). However, it should be pointed out that strategies based on shrinkage estimators tend to better perform during and immediately after the sovereign debt crisis. \\
Among the ERC portfolios (see Figure  \ref{fig:BI2Y_ERC}), we have a remarkable prevalence of network-based approaches. Also in this case, sample estimators assure the best performance.  According to  the GMV portfolio (see Figure \ref{fig:BI2y_GMV}), the four alternative strategies show a similar pattern, with a slight preponderance of the NB-shrinkage approach. Notice that in Figure \ref{fig:BI2y_GMV} the performance of the NB-sample (the blue line) is not visible, as it is overlapped with that of the NB-shrinkage. \\
Figure \ref{fig:BI2Y_Best_Bench} collects the best out-of-sample performances for each \textit{risk based} analysed portfolio (i.e. MDP, ERC and GMV). What emerges is that the network-based approaches outperform classical strategies and the highest out-of-sample performance at the end of the period is assured by the NB-sample MDP portfolio. However, while MDP and ERC 
have a similar behaviour over time, the NB-sample GMV strategy shows a lower volatility, performing better during financial turbulence periods.

Concluding, from the analysis of the performances of Smart Based portfolios with the various approaches, the network-based models almost always lead to higher out-of-sample performance compared to the corresponding classical ones. In particular, NB-sample MDP is the best portfolio. This result is confirmed also by the analysis  carried  on S\&P dataset, for both estimators of the covariance matrix, i.e the sample or the shrinkage towards the constant correlation approach, and for all the rolling windows strategies. \\
However, the simple inspection of the Figure \ref{fig:smart_beta_perf} is not enough in identifying  the best portfolio selection strategy. To this end,  in order to complete the analysis
we report in Table \ref{T:stat_bench} the four moments of the out-of-sample returns' distributions and alternative performance measures (namely, $SR$, $IR(EW)$\footnote{The $IR(EW)$ indicates the Information ratio where the reference portfolio is the Equally Weighted one.} and  $OR$).  As well-known, these performance measures consider different characteristics of the portfolios and they could may lead to different rankings  between the models. However, by the inspection of Table \ref{T:stat_bench}, we can provide additional insights.

\begin{table}[H]
	\centering
	{\footnotesize \resizebox{0.6\textwidth}{!}{
			\begin{tabular}{lccccc}
				\hline
				\multicolumn{6}{c}{ERC} \\
				\hline \hline
				& S-sample & S-shrinkage & NB-sample & NB-shrinkage & EW     \\
				$\mu^{\star}$       &  7.53E-05 &   7.12E-05 &   1.15E-04 &   7.30E-05 &   6.19E-05  \\
				$\sigma^{\star}$    & 0.005    & 0.005       & \textbf{0.004}      & \textbf{0.004}        & 0.008  \\
				$skew^{\star}$      & -1.157   & -1.185      & \textbf{-0.793 }    & -1.361       & \textbf{-0.725} \\
				$kurt^{\star}$      & 20.376   & 20.673      & \textbf{15.385}     & \textbf{19.219}       & 22.606 \\
				$SR^{\star}$        & 0.014    & 0.013       &\textbf{ 0.027}      &\textbf{ 0.017  }      & 0.007  \\
				$IR^{\star}(EW)$    & 0.003    & 0.002       & \textbf{0.010}      & \textbf{0.004}        &        \\
				$OR^{\star}$ & 1.046    & 1.044       & \textbf{1.085}      & \textbf{1.054}        & 1.026  \\
				\hline
				\multicolumn{6}{c}{MDP} \\
				\hline \hline
				& S-sample & S-shrinkage & NB-sample & NB-shrinkage & EW     \\
				$\mu^{\star}$       &  1.25E-04 &   7.55E-05 &   1.34E-04 &   1.17E-04 &   6.19E-05   \\
				$\sigma^{\star}$    & 0.006    & \textbf{0.004}       & 0.006      & 0.007        & 0.008  \\
				$skew^{\star}$      & -3.587   & -1.146      & \textbf{0.440}      & \textbf{0.366}        & -0.725 \\
				$kurt^{\star}$      & 77.328   & 44.393      & \textbf{17.706}     & \textbf{24.970}       & 22.606 \\
				$SR^{\star}$        & \textbf{0.021 }   & 0.018       & \textbf{0.021}      & 0.017        & 0.007  \\
				$IR^{\star}(EW)$    & 0.008    & 0.002       & \textbf{0.008}      & \textbf{0.009}        &        \\
				$OR^{\star}$ & 1.068    & 1.063       & \textbf{1.071}      &\textbf{ 1.070}        & 1.026  \\
				\hline
				\multicolumn{6}{c}{GMV} \\
				\hline \hline
				& S-sample & S-shrinkage & NB-sample & NB-shrinkage & EW     \\
				$\mu^{\star}$       & 6.07E-05 &   6.14E-05 &   6.03E-05 &   6.05E-05 &   6.19E-05   \\
				$\sigma^{\star}$    & 0.0018    & 0.0018       & 0.0017      &\textbf{ 0.0016}        & \textbf{0.008}  \\
				$skew^{\star}$      & \textbf{-0.448}   & \textbf{-0.363}      & -0.517   & -0.504       & -0.725 \\
				$kurt^{\star}$      & 9.875    & 9.686       & \textbf{8.569}      & \textbf{8.525}        & 22.606 \\
				$SR^{\star}$        & 0.033    & 0.034       & \textbf{0.035 }     & \textbf{0.036 }       & 0.007  \\
				$IR^{\star}(EW)$    & 0.000    & 0.000       & 0.000      & 0.000        &        \\
				$OR^{\star}$ & \textbf{1.100}    & \textbf{1.102}      & 1.093     & 1.093        & 1.026 \\
				\hline			
			\end{tabular}  	
	}}
	\caption{
		Out-of-sample statistics for the \textit{risk-based} approaches in case of Banks \& Insurers portfolio with a buy and hold strategy of 2 years in-sample and 1 month out-of-sample.		For each strategy (ERC, MDP and GMV), both sample and shrinkage estimators are reported for classical and network-based models. The last column also considers results for EW. The two best results are reported in bold for each measure. 	
		All the statistics are reported on daily bases. }
	\label{T:stat_bench}
\end{table}

First, we observe that for both ERC and GMV strategy, the network-based approach leads to lower out-of-sample risk, measured by the standard deviation, regardless of the  estimation method of the covariance matrix. Moreover, for each strategy the network-based approach (using  either sample or shrinkage estimator), almost always leads to higher skewness and lower kurtosis with respect to the corresponding standard approach. These findings are further confirmed by $SR$ and $OR$ values.  In particular, it results that for both the ERC and MDP strategies, the best portfolio is obtained thorough the NB-sample approach, because to this correspond the best values of all parameters.  
As regard to the GMV strategy,  the best portfolio is the NB-shrinkage  approach. Hence, the results reported in Table \ref{T:stat_bench} make us more confident in believing that using the network theory in building the \textit{smart beta} strategies  can be a good alternative to the standard approach not only for the easy visualization of the  results (as reported in Figures \ref{F:BI2Ynet} and \ref{fig:smart_beta_perf}) but also for the better performances that they may reach in an out-of-sample perspective. 
The conclusions drawn for Banks and Insurers portfolio with a  monthly stepped two-years rolling window are still valid, in general, also for the other rolling window under analysis (namely six months in-sample  and one month out-of-sample) and for the S\&P portfolio\footnote{Detailed results are reported in the Supplementary material}. 

Let us now analyse the results obtained in case of the Mean-Variance portfolio where different levels of trade-off between risk and return are considered. In particular we report the results for $\lambda$ equal to $0.2, \ 0.4,\ 0.6, \ 0.8$, respectively. A low level of $\lambda$ indicates that the investor   gives higher importance to the portfolio return. In particular, $\lambda=0$ indicates that the  decision maker is completely ignoring the risk of the portfolio. In this  case, the optimal portfolio  is usually  concentrated only in the asset with the higher return\footnote{The optimal portfolio when $\lambda=0$ may not be unique, since more than one asset with the highest return can exist.}. On the contrary, high values of $\lambda$ indicate a higher relevance to the risk with respect to the return. The extreme case of   $\lambda=1$ corresponds to the GMV portfolio\footnote{These results are reported in Table \ref{T:stat_bench}}, meaning that the decision maker completely ignores the portfolio return.

\begin{figure}[H]
	\centering
	\subfloat[$\lambda=0.2$]{\fbox{\includegraphics[ height=5cm, width=7.5cm]{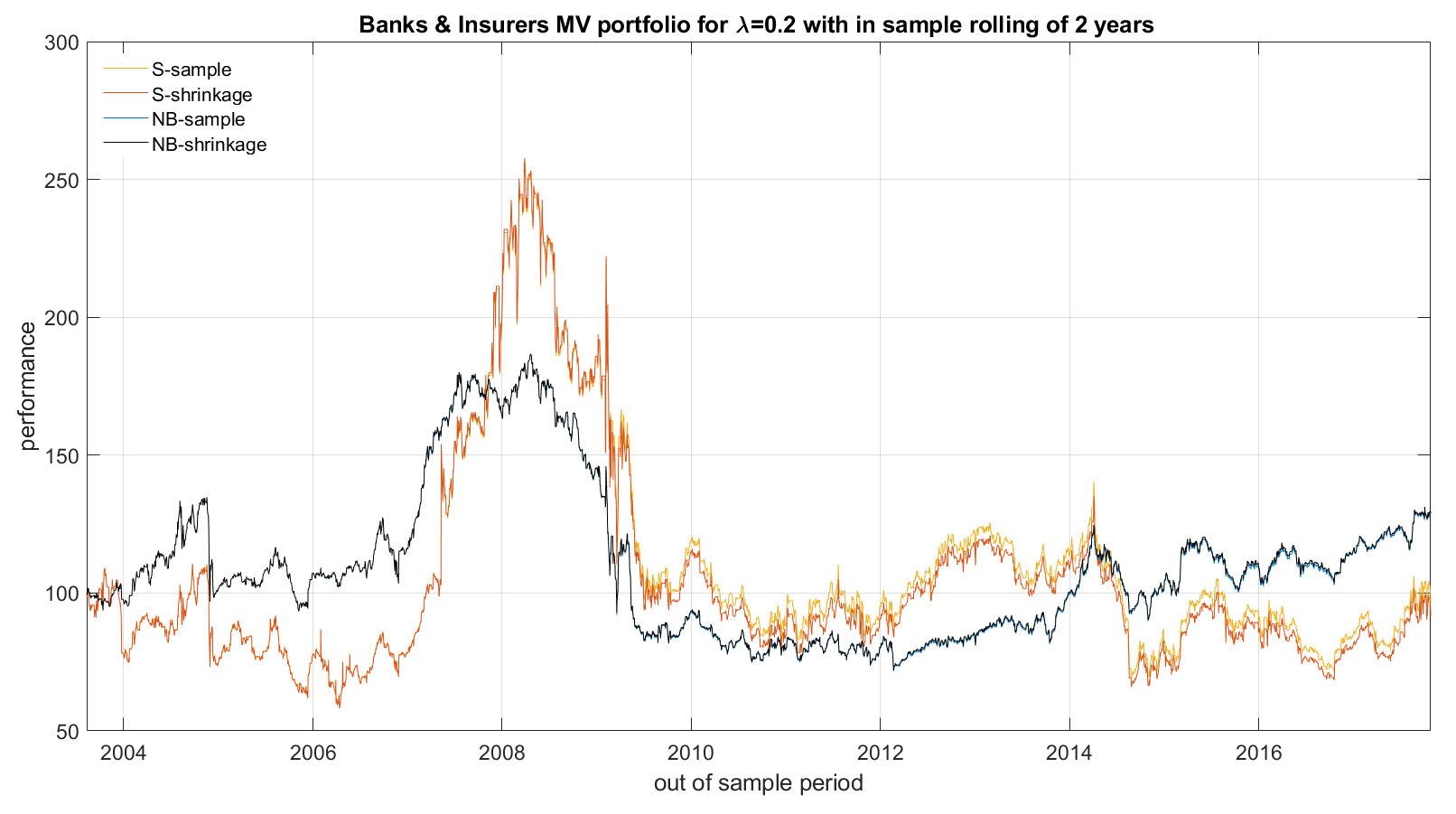}}\label{fig:BI2Y_MVLam02}}\ \hspace{0.5mm}
	\subfloat[$\lambda=0.4$]{\fbox{\includegraphics[ height=5cm, width=7.5cm]{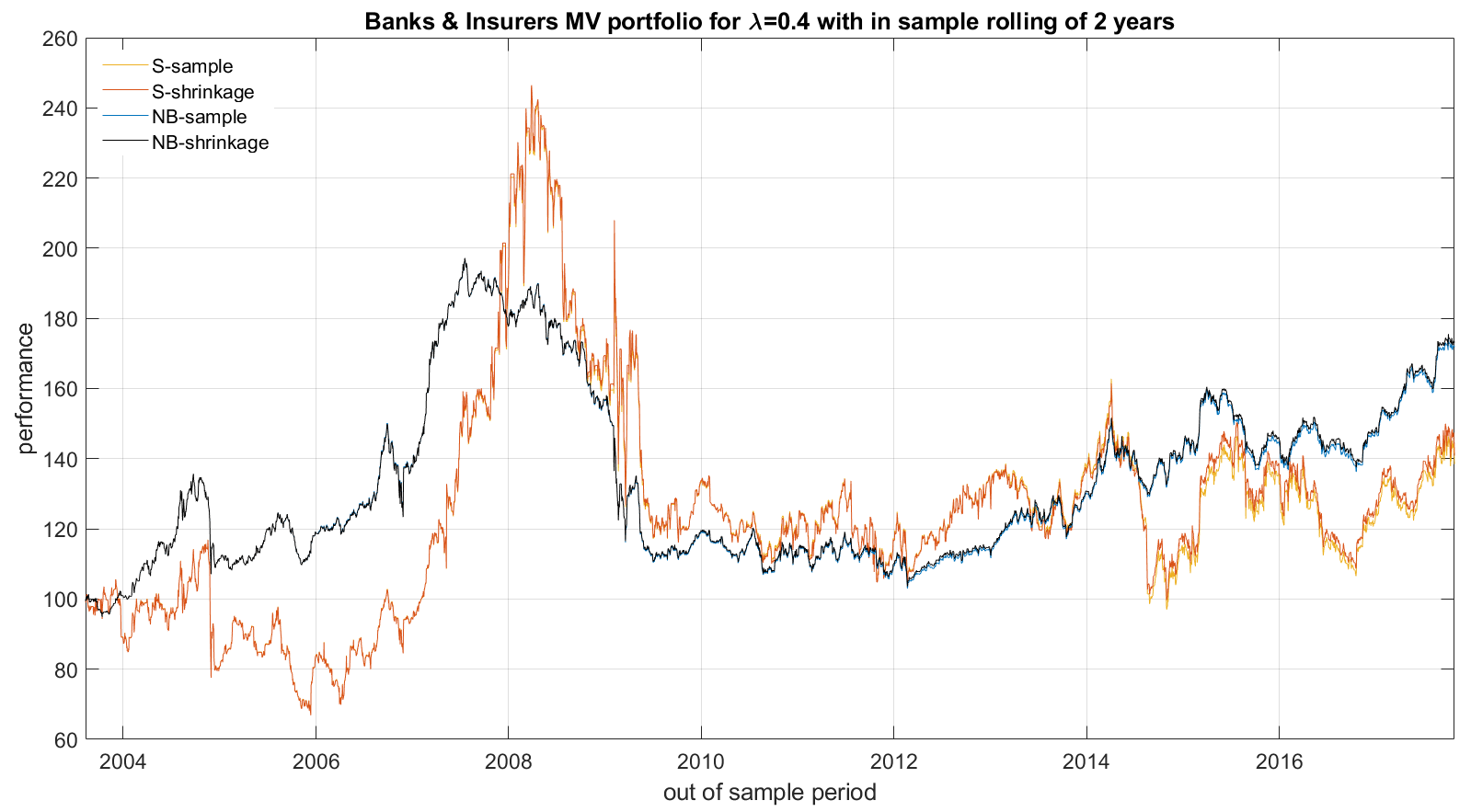}}\label{fig:BI2Y_MVLam04}}\
	\subfloat[$\lambda=0.6$]{\fbox{\includegraphics[ height=5cm, width=7.5cm]{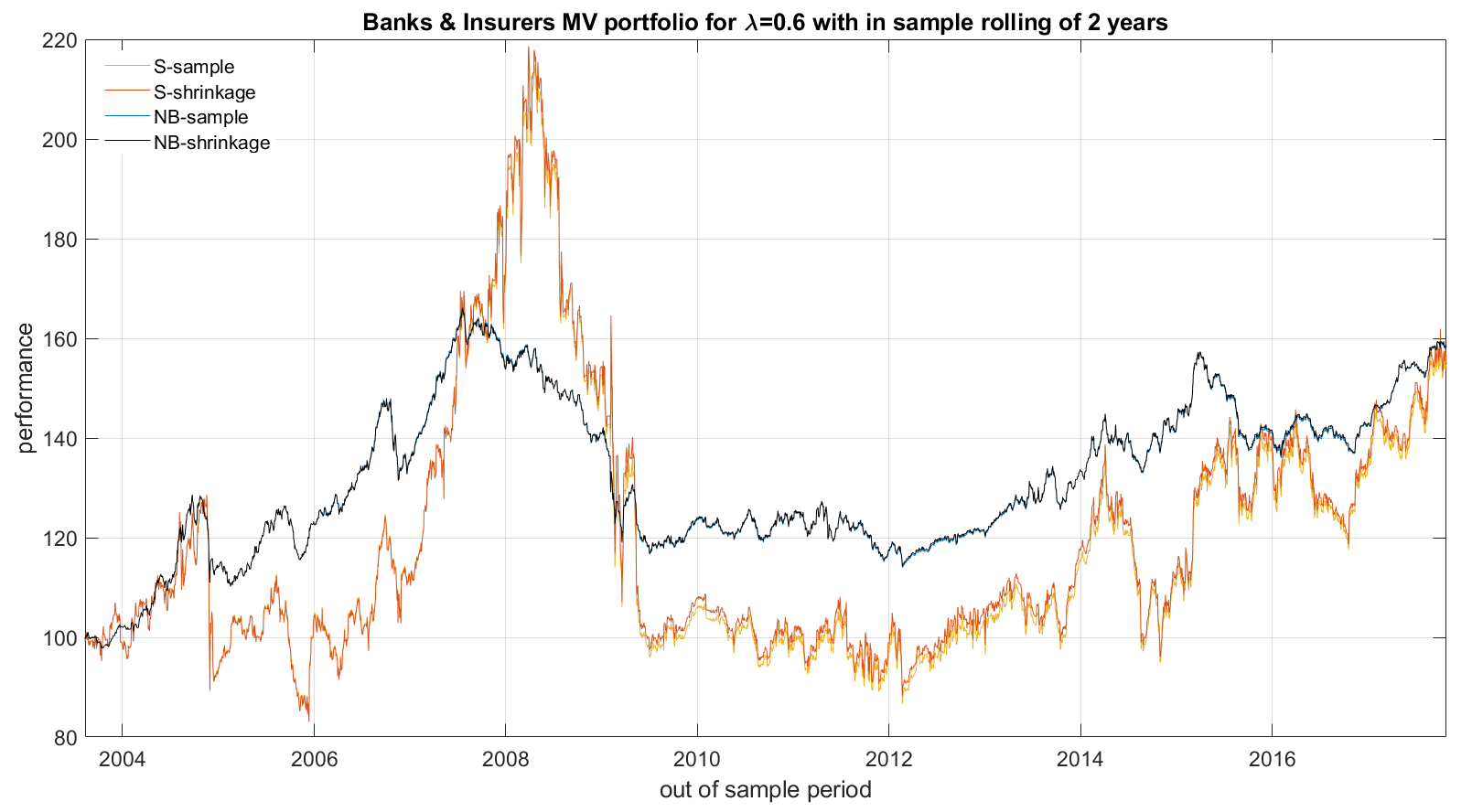}}\label{fig:BI2Y_MVLam06}}\ \hspace{0.5mm}
	\subfloat[$\lambda=0.8$]{\fbox{\includegraphics[ height=5cm, width=7.5cm]{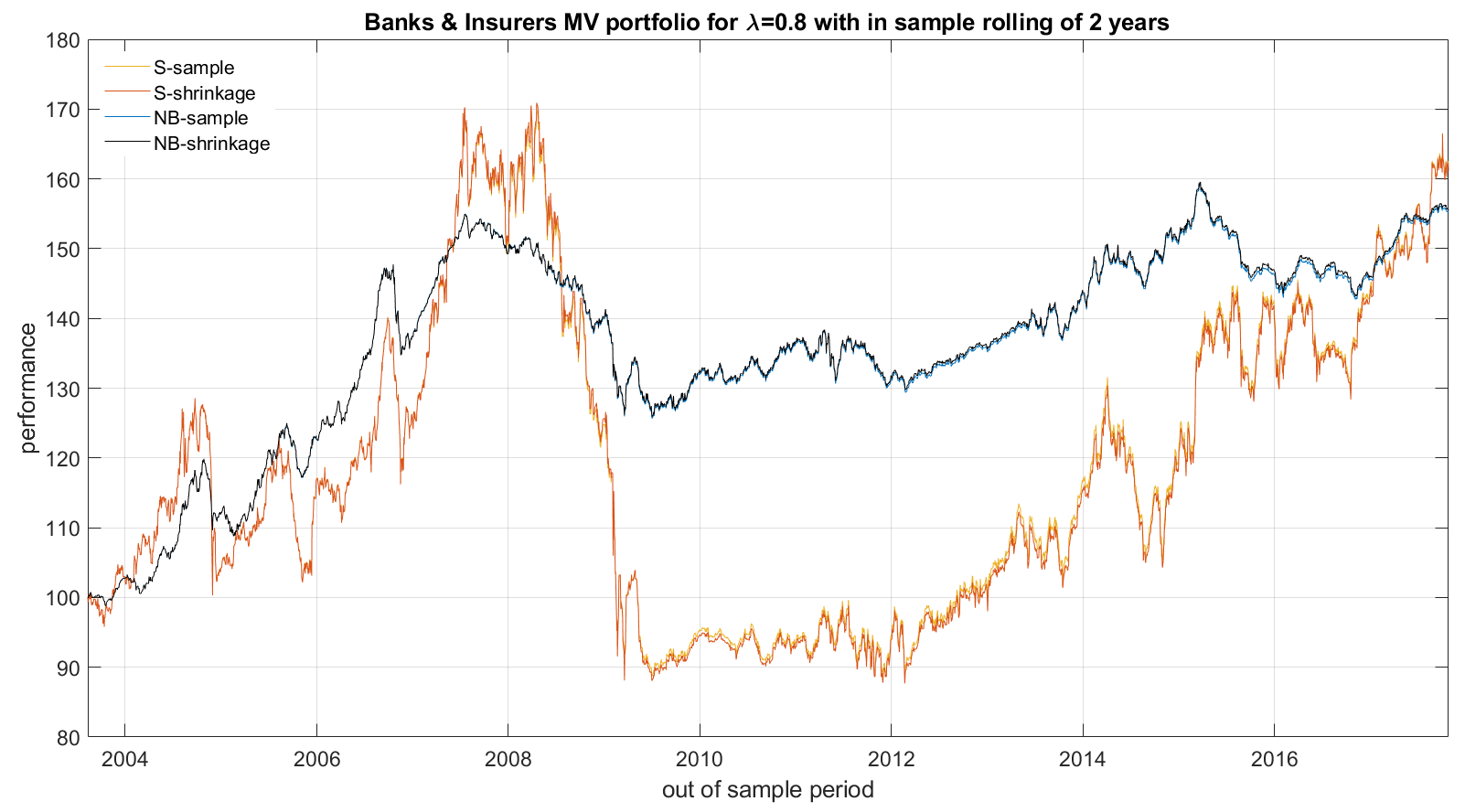}}\label{fig:BI2Y_MVLam08}}
	\caption{Out-of-sample performances for Banks and Insurers portfolio with a rolling window of  2 years in-sample and 1 month out-sample. In Figures \ref{fig:BI2Y_MVLam02}, \ref{fig:BI2Y_MVLam04}, \ref{fig:BI2Y_MVLam06} and \ref{fig:BI2Y_MVLam08}   we report the out-of-sample performances for S-sample, S-Shrinkage, NB-sample and NB-Shrinkage strategies according to alternative values of the trade-off parameter (namely, $\lambda=0.2$, $\lambda=0.4$, $\lambda=0.6$ and $\lambda=0.8$ respectively). }
	\label{fig:MV_perfor}
\end{figure}
\noindent
Figure \ref{fig:MV_perfor} reports the out-of-sample performances of the Banks and Insurers dataset, with a buy and hold strategy of two years in-sample and one month out-of-sample, in case of the MV model. As previously described in subsection \ref{estMeth}, for this strategy $\bm{\mu}$ and $\bm{\Sigma}$ have to be estimated. We
estimate $\bm{\mu}$ using the sample approach while $\bm{\Sigma}$ is estimated using both the sample and the shrinkage toward constant correlation methods.  Notice that the matrix $\bm{\Sigma}$ is also used in the network-based approaches to construct the network and to obtain the interconnectedness matrix $\bm{C}$.  Figure \ref{fig:BI2Y_MVLam02}  displays  the out-of-sample performances obtained setting $\lambda=0.2$, which means that  the investor tends to prefer high potential returns with respect to low levels of uncertainty. 
Although it is not possible to define a univocal ranking between methods in terms of performance, we observe higher returns at the end of the period with the network-based approaches. In this case standard methods behave better in periods of crisis. 
This fact is partially confirmed also for other values of the trade-off parameters, as reported in Figures \ref{fig:BI2Y_MVLam04} (for $\lambda=0.4$),  \ref{fig:BI2Y_MVLam06} (for $\lambda=0.6$) and  \ref{fig:BI2Y_MVLam08} (for $\lambda=0.8$). In particular, it is noticeable the prevalence of NB models after the crisis of 2008 for $\lambda=0.8$.  

To have a complete view  of the effect of the network based strategy on the MV model, we report in Table \ref{tab:statts_MV_BI_2Y} the first four moments  and alternative performance measures. Notice that, in this case, the $IR(S-sample)$ indicates the Information ratio computed taking as a reference the standard sample approach optimal portfolio. 

\begin{table}[H]
	{\footnotesize \resizebox{1\textwidth}{!}{
			\centering
			\begin{tabular}{lcccc|cccc}
				\hline
				\multicolumn{5}{c}{$\lambda=0.2 $} &  \multicolumn{4}{c}{$\lambda=0.4 $} \\
				\hline \hline
				& S-sample  & S-shrinkage  & NB-sample & NB-shrinkage  & S-sample  & S-shrinkage & NB- sample & NB-shrinkage\\
				$\mu^{\star}$  & {\bf 2.61E-04} & {\bf 2.58E-04} &   1.35E-04 &   1.35E-04 & {\bf 2.42E-04} & {\bf 2.49E-04} &   1.77E-04 &   1.79E-04\\
				$\sigma^{\star}$  & 0.024 & 0.024 & \textbf{0.012} & \textbf{0.012} & 0.017 & 0.018 & \textbf{0.007} & \textbf{0.007} \\
				$skew^{\star}$    & \textbf{3.566} & \textbf{3.814} & -0.586 & -0.581   & \textbf{0.521} & \textbf{0.561} & -0.425 & -0.432\\
				$kurt^{\star}$  & 85.515 & 92.514 & \textbf{17.888} & \textbf{17.814} & 22.904 & 22.868 & \textbf{15.476} & \textbf{15.625} \\
				$SR^{\star}$    & 0.011 & 0.011 & \textbf{0.012} & \textbf{0.012}  & 0.014 & 0.014 & \textbf{0.024} & \textbf{0.024} \\
				$IR^{\star}(S-sample)$   &       & -0.003 & -0.007 & -0.007   &       & 0.011 & -0.003 & -0.003 \\
				$OR^{\star}$ & \textbf{1.040} & \textbf{1.040} & 1.038 & 1.038 & 1.047 & 1.048 & \textbf{1.076} & \textbf{1.077} \\
				\hline
				\multicolumn{5}{c}{$\lambda=0.6$} &   \multicolumn{4}{c}{$\lambda=0.8$} \\
				\hline \hline
				& S-sample  & S-shrinkage &NB-sample & NB-shrinkage  & S-sample  & S-shrinkage &NB- sample & NB-shrinkage\\
				$\mu^{\star}$  & {\bf 2.05E-04} & {\bf 2.11E-04} &   1.36E-04 &   1.36E-04 & {\bf 1.68E-04} & {\bf 1.69E-04} &   1.25E-04 &   1.26E-04 \\
				$\sigma^{\star}$  & 0.013 & 0.013 & \textbf{0.004} & \textbf{0.004}& 0.008 & 0.009 & \textbf{0.003} & \textbf{0.003} \\
				$skew^{\star}$    & \textbf{-0.195} & \textbf{-0.142} & -0.545 & -0.549   & -0.489 & -0.507 & \textbf{-0.443} & \textbf{-0.439} \\
				$kurt^{\star}$  & 19.018 & 19.136 & \textbf{10.242} & \textbf{10.365} & 13.960 & 14.413 & \textbf{7.924} & \textbf{7.964} \\
				$SR^{\star}$    & 0.016 & 0.016 & \textbf{0.032} & \textbf{0.032} & 0.020 & 0.020 & \textbf{0.046} & \textbf{0.046}\\
				$IR^{\star}(S-sample)$   &       & 0.010 & -0.006 & -0.006  &       & 0.004 & -0.006 & -0.006\\
				$OR^{\star}$ & 1.051 & 1.051 & \textbf{1.074} & \textbf{1.076} & 1.062 & 1.061 & \textbf{1.085} & \textbf{1.083}  \\
				\hline
			\end{tabular}%
	}}
	\caption{	Out-of-sample statistics for the MV model, in case of the Banks \& Insurers portfolio with 2 years rolling estimation windows for the mean vector and the covariance matrix and one month for out-of-sample returns. The two best results are reported in bold for each measure. All the statistics are reported on daily bases. }
	\label{tab:statts_MV_BI_2Y}%
\end{table}%

\noindent
The results in Table \ref{tab:statts_MV_BI_2Y} clearly shows that, for all considered values of $\lambda$,
the MV network-based portfolio has lower risk (measured by the standard deviation) than the corresponding standard approach. Moreover, in general the network-based portfolios lead also to higher out-of-sample performances in terms of Sharpe ratio and Omega ratio.  These results are in line with those obtained for the \textit{risk-based} approaches (presented in Figure \ref{fig:smart_beta_perf} and Table \ref{T:stat_bench}) enforcing our believe that applying network tools to portfolio selection models may enhance the portfolio selection process. \\
At the end, we observe that the use of network tools to manage the optimal portfolio selection is effective, especially in the case of \textsl{risk-based} strategies. Looking at the results of the MV portfolios, in the supplementary material we can observe that for some levels of the trade-off parameters the network-based portfolios lead to better out-of-sample results, while for other values of the trade-of parameter the standard approaches behave better.  It is not clear for the moment how the trade-off parameter $\lambda$ influence the results obtained. We believe this is also related to the estimator used for the mean vector.


\section{Conclusions} \label{concl}
In this work we applied networks tools to the most used portfolio models characterized by an objective function depending on the covariance matrix of assets. 
Following \cite{ANORclegrahit}, we took advantage of the correlations network to capture the interconnectedness between assets, that explicitly enters through the clustering coefficient in the objective function. 
We extended the approach of \cite{ANORclegrahit}, tested to the GMV model, proposing the application of network theory also for the most used Smart-Beta models, as well as the MV model.
We estimated the network-correlation structure through the sample approach (as in \cite{ANORclegrahit}) and through the shrinkage toward the constant correlation.

To test the robustness of our methodology,  we performed numerical analyses, based on two large-dimensional portfolios.
We implemented both the standard and the network-based models, using sample and shrinkage estimators for the covariance matrix, and we compared the out-of-sample performances based on a rolling sample optimization. 
The results obtained show in most cases the effectiveness of network-based portfolios compared to the standard approaches and to the equally weighted portfolios. Network-based strategies show higher out-of-sample performances and lower out-of-sample volatility, reducing the risk.
Results appeared significant especially for Smart-Beta strategies, which are based only on the risk measure, that refers to the covariance matrix or the interconnectedness matrix. 
On the contrary, results in case of mean-variance portfolio do not provide a univocal ranking of the models.  Indeed, the network-based approaches lead to better results in the most cases, but there is a good percentage of cases in which the standard approaches  behave  better.  This behaviour depends obviously  on the  trade-off parameter value and to the estimator used for the portfolio mean.  We believe that better out-of-sample results can be obtained in case the network theory is used for the estimation of not only the risk measure but also for the estimation of the performance measure, which is left for future research. \\
Concluding, we hope that this empirical analysis will help to shed some light on how network theory can be implemented in  portfolio  selection problems and encourage portfolio managers in considering and testing the network-based portfolio selection models as an alternative to the standard approaches. 


\bibliographystyle{spmpsci}      

\end{document}